# A liquid implant that grows into a stable scaffold *in situ*

Authors: Wenqiang Li[1#], Qichen Peng[1#], Fuhong Wang[2#], Zhentan Quan[2], Shengxiang Cai[2], Zhuoming Zhou[3], Jin Zhang[2], Ting Liu[1], Xuanping Xie[1], Lin Zhu[4], Bo Wang[2], Shuang Wang[1], Yuan An[1], Yingtong Zhou[2], Zirong Liu[2], Haohui Li[2], Songshen Li[2], Ben Wang[5], Xinjian Yang[1*], Quanshui Zheng[2], Hongzhang Wang[2*]

**Affiliations:**

[1]Department of Neurosurgery, Beijing Tiantan Hospital and Beijing Neurosurgical Institute, Capital Medical University, Beijing, 100071, P. R. China.

[2]Institute of Materials Research, Center of Double Helix, Shenzhen International Graduate School, Tsinghua University, Shenzhen 518055, P. R. China.

[3]Center for Nanomedicine and Department of Anesthesiology, Brigham and Women's Hospital, Harvard Medical School, Boston, MA, USA.

[4]Department of Oral and Maxillofacial-Head and Neck Oncology, Beijing Stomatological Hospital, Capital Medical University, Beijing 100050, China

[5]College of Chemistry and Environmental Engineering, Shenzhen University, Shenzhen 518055, P. R. China.

#These authors contributed equally to this work.

*Corresponding authors.

Email: wanghz@sz.tsinghua.edu.cn (H.W); yangxinjian@voiceoftiantan.org (X.Y)

**Abstract:** Endovascular embolization requires materials that can pass through narrow microcatheters yet form stable, conformal occlusions within irregular vascular lesions, but reconciling these competing requirements remains challenging. Here we report a wirelessly activated liquid precursor whose final architecture is constructed only after target localization. A gallium-based liquid metal containing iron microparticles and gas microcavities remains deformable, magnetically navigable, and retrievable before activation. Wireless induction heating activates the internal reactive microenvironment, coupling gas-driven microcavity growth with oxidative mineralization of newly generated gallium interfaces to form a porous, self-supporting scaffold. In aneurysm and arteriovenous malformation models, activation enabled durable occlusion, aneurysm-neck healing, and six-month stability in a porcine aneurysm model. These findings establish post-delivery construction as a strategy for creating stable therapeutic implants from minimally invasive liquid precursors.

**Main Text:**

The ability to reconstruct diseased vasculature through minimally invasive delivery has transformed modern endovascular therapy (*1*-*3*). Endovascular therapy requires therapeutic materials to traverse narrow microcatheters and confined, tortuous vasculature before occupying lesions that are substantially larger, geometrically irregular, and continuously exposed to blood flow. This challenge is particularly evident in intracranial aneurysms, localized dilations of cerebral arteries whose rupture can cause catastrophic subarachnoid hemorrhage (*4*-*6*). Effective embolization requires a material to conformally occupy the aneurysm sac while maintaining stable exclusion from the circulation over time (*7*). Incomplete filling or insufficient fixation can leave residual flow, permit material displacement, and ultimately lead to recanalization. A central challenge is therefore to reconcile the small material footprint required for endovascular delivery with the larger therapeutic volume and persistent mechanical stability required after implantation (*8, 9*).

Current embolic technologies address different aspects of this challenge but do not eliminate the underlying mismatch (*2, 10*). Detachable coils provide a mechanically stable framework but fill vascular lesions through discrete packing; incomplete occlusion and subsequent recurrence remain limitations (*11*). Liquid embolics and other stimuli-responsive materials take a complementary approach, conforming to irregular vascular spaces before acquiring structural stability through precipitation, gelation, polymerization, swelling, or shape recovery. Yet before fixation is established, these aqueous or solvent-based precursors remain vulnerable to hemodynamic washout and nontarget migration, particularly in high-flow lesions (*12*). Moreover, because structural transformation typically begins during or soon after deployment, delivery and fixation remain tightly coupled, leaving only a limited window for repositioning or retrieval.

Stimuli-responsive gallium-based liquid metals have enabled functions ranging from electrical circuit restoration to magnetically controlled manipulation and mechanically adaptive devices (*13*-*17*). A different strategy for endovascular implantation would be to exploit this material adaptability while delaying structural commitment until target localization. Inspired by biological growth, where transportable precursors generate stable structures only after reaching their destination, we reasoned that an endovascular implant could likewise remain deformable and maneuverable during delivery, yet grow into its load-bearing scaffold only after remote activation. Rather than simply solidifying or swelling, the liquid would form and stabilize a new porous architecture as it expands, thereby decoupling the small material footprint required for endovascular transport from the larger functional volume and mechanical stability required for therapy (*18, 19*).

Here we report a wirelessly activated growing liquid implant that realizes this design principle. The implant is based on a gallium liquid-metal precursor containing iron microparticles and prestructured internal microcavities. Before activation, it remains cohesive, deformable, magnetically controllable, and retrievable under flow. After target localization, wireless induction heating triggers water-mediated interfacial reactions, coupling gas-driven cavity growth with oxidative mineralization of newly generated interfaces to form a larger, self-supporting porous scaffold. This delayed structural construction allows the material to remain controllable during delivery while generating its final therapeutic volume and architecture only at the target site. In aneurysm models, the implant enabled controlled delivery, durable occlusion, progressive neck healing, and long-term stability in porcine aneurysm model; the strategy was further extended to a large-animal model of arteriovenous malformation. Together, these results establish postdelivery scaffold construction from a controllable liquid precursor as a strategy for endovascular implantation.

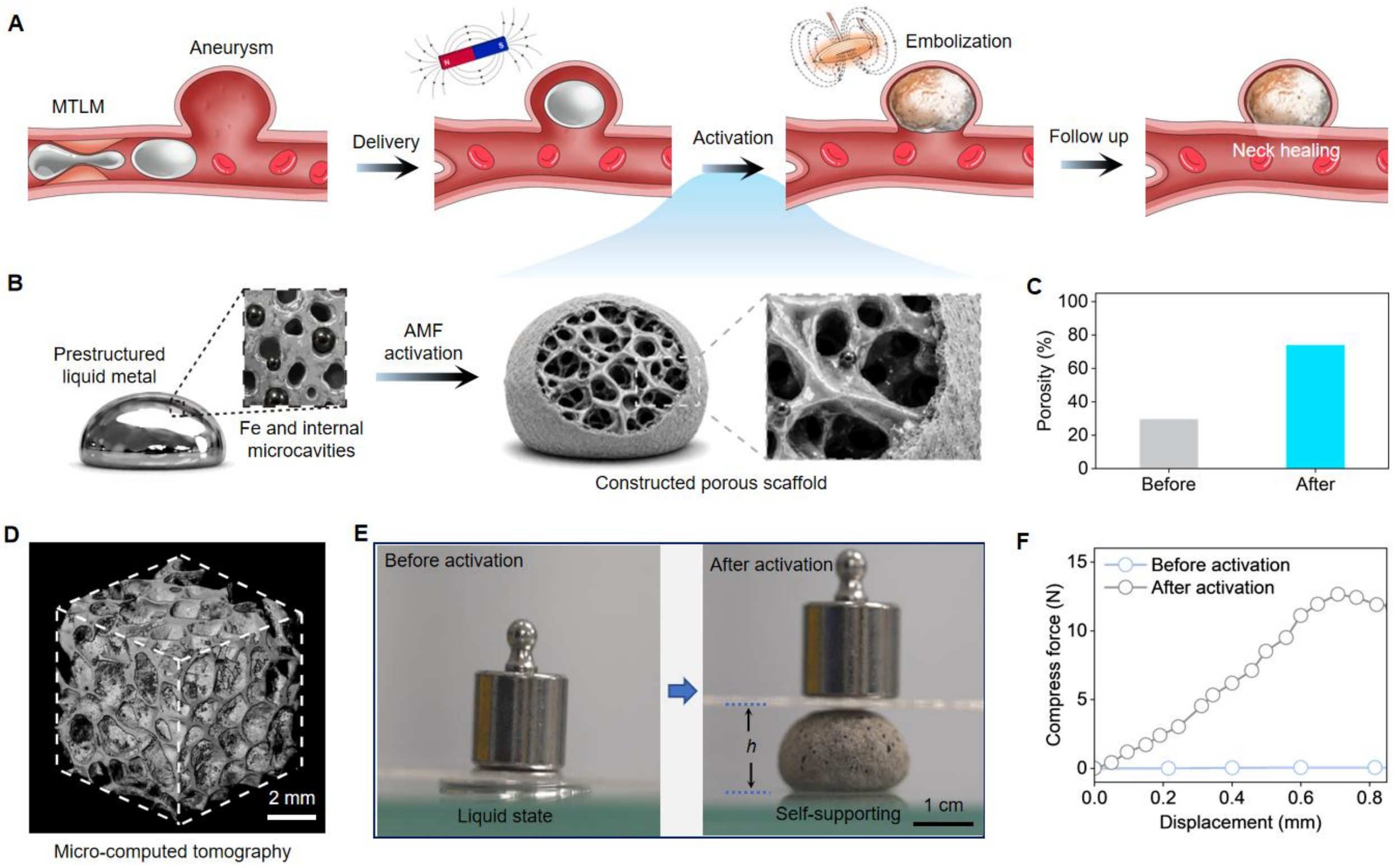

**Fig. 1. Remote triggering enables MTLM to undergo a delivery-to-fixation transition.**

(**A**) Schematic illustration of the magnetothermal liquid-metal implant (MTLM) embolization strategy. (**B**) Schematic of the transformation of MTLM from a liquid-metal state containing preconfigured cavities into a porous scaffold after alternating magnetic field (AMF) activation. (**C**) Porosity of MTLM before and after AMF activation. (**D**) Three-dimensional CT reconstruction of activated MTLM, showing a porous architecture distributed throughout the material. (**E**) Photographs of the macroscopic load-bearing capacity of MTLM before and after activation. (**F**) Load–displacement curves of MTLM before and after AMF activation.

## A growing liquid precursor constructs a load-bearing scaffold after delivery

We developed a MTLM with distinct material states for vascular transport and post implantation function. MTLM can be delivered in a deformable liquid state and guided into the aneurysm sac under an external magnetic field. Subsequent activation by an AMF induces volumetric expansion and structural fixation, resulting in aneurysm occlusion that promote gradual healing of the aneurysm neck (Fig. 1A). MTLM consists of a eutectic Ga-In (EGaIn) liquid-metal precursor containing Fe microparticles and internal microcavities. Before AMF activation, it retains its fluidity, deformability, and magnetic responsiveness. Owing to these properties, MTLM can be delivered through a microcatheter or guided by an external magnetic field. Its deformability also allows it to conform to the irregular geometries of blood vessels and aneurysm sacs. Exposure to an AMF subsequently triggers volumetric growth and formation of a mechanically persistent porous structure. Thus, MTLM is delivered as a deformable liquid precursor and develops its load-bearing architecture only after deployment.

Before activation, MTLM retained liquid-like reconfigurability. Separately dispensed droplets readily merged into a continuous body, and the material conformally occupied cavities of different geometries without requiring a predefined implant shape (fig. S1). AMF activation produced a markedly different state: the compact precursor expanded substantially and developed a three-dimensional porous structure (Fig. 1B), along with porosity increase (Fig. 1C). CT images further showed that the porous structure extended throughout the bulk rather than being limited to the surface (Fig. 1D). AMF activation thus increased the material volume and generated an internal porous structure, rather than merely fixing the precursor's initial geometry.

Volumetric growth was accompanied by the emergence of load-bearing function. Before activation, MTLM readily deformed under external loading, consistent with the liquid metal. After AMF activation, the expanded architecture retained its geometry under substantially greater loads (fig. S2). We applied external loads to MTLM before and after AMF activation (Fig. 1E). The AMF-activated MTLM formed a self-supporting structure, whereas the liquid MTLM underwent substantial deformation under loading. In addition, compression tests demonstrated the enhanced compressive resistance of the AMF-activated MTLM (Fig. 1F). Therefore, activation generated not only additional volume and porosity, but also the mechanical stability required for the expanded architecture to persist as an implant. These results establish MTLM as a deformable liquid precursor that develops a larger, porous, and load-bearing architecture only after remote activation.

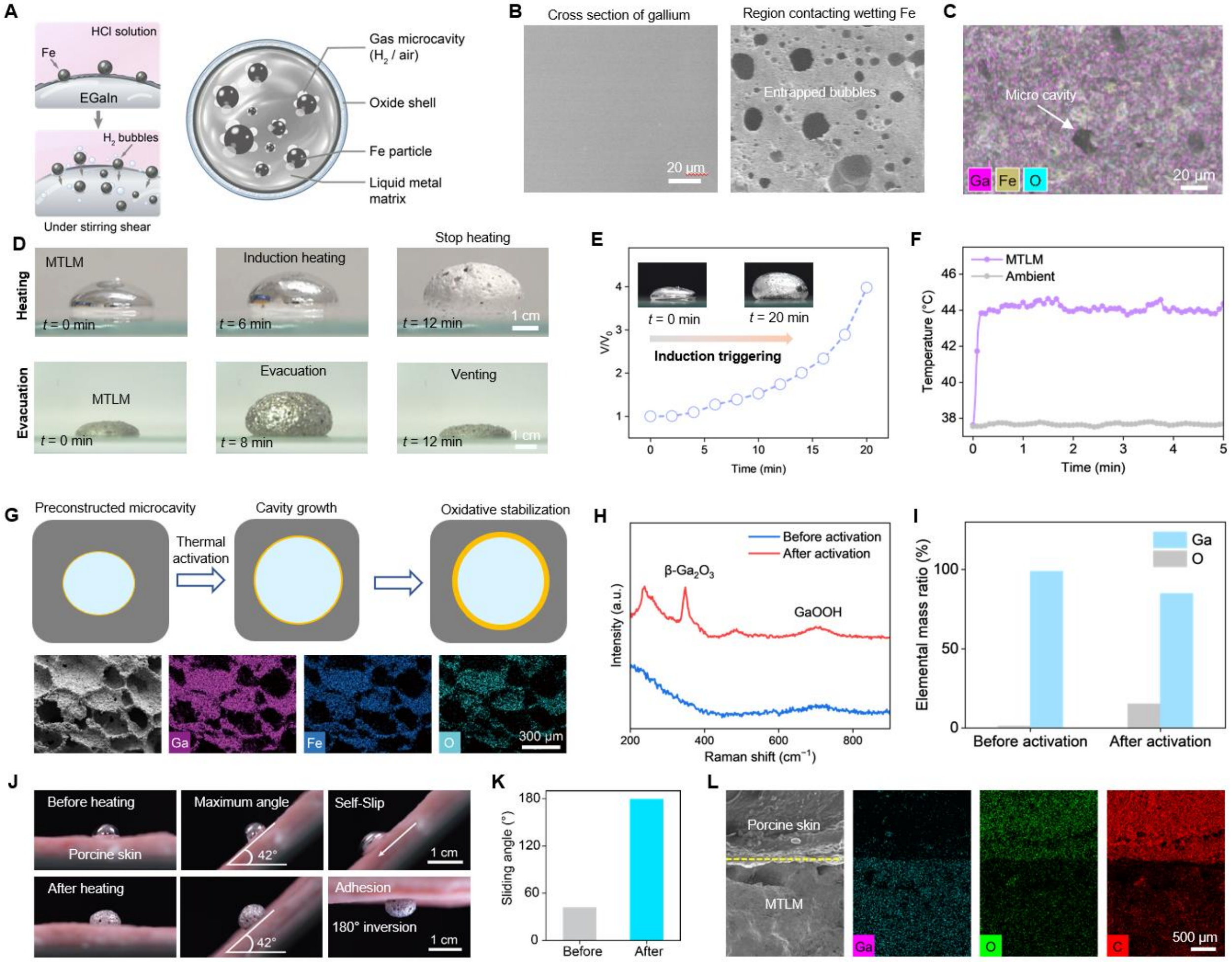

**Fig. 2. Microcavity expansion and oxidative locking cooperatively drive the transformation of MTLM from a liquid into a porous scaffold.**

(**A**) Schematic illustration of the hierarchical internal structure of MTLM, comprising a liquid-metal matrix, Fe particles, oxide-stabilized gas and microcavities. (**B**) Representative images of cross-sections of pure Ga and MTLM. (**C**) The distributions of Ga, Fe and O around a representative microcavity in MTLM. (**D**) Representative images comparing AMF-induced expansion with pressure-induced deformation of MTLM. (**E**) Time-dependent volumetric expansion ratio $V/V_0$ of MTLM during AMF activation. Insets show representative material morphologies at different time points. (**F**) Temperature changes of MTLM and the surrounding environment during AMF activation. (**G**) Schematic of the expansion-coupled oxidative-locking mechanism. Elemental maps below show the spatial distributions of Ga, Fe, and O in activated MTLM. (**H**) Raman spectra of MTLM before and after activation. (**I**) EDS-derived relative mass fractions of Ga and O in MTLM before and after AMF activation, showing an increased oxygen fraction after activation. (**J**) Representative images of the adhesion test of MTLM on porcine skin before and after activation. Nonactivated MTLM showed limited attachment to the tissue surface, whereas activated MTLM remained attached during tissue inversion. (**K**) Quantification of the sliding angle of MTLM before and after activation, showing an increase from approximately 42° to 180° following activation. (**L**) Cross-sectional electron micrograph and elemental maps of the interface between activated MTLM and tissue.

## Expansion-templated mineralization generates a persistent porous scaffold

MTLM was prepared by mixing EGaIn with Fe microparticles in an acidic environment, where acid-assisted wetting of Fe promoted its incorporation into the liquid metal (*20*). During acid-assisted wetting, gas evolved as EGaIn incorporated Fe microparticles, resulting in the formation of microcavities (Fig. 2A). Subsequent washing and vacuum drying removed residual acid and excess free water (fig. S3). The resulting stable liquid precursor retained its prestructured internal microcavities without spontaneous expansion when exposed to air (fig. S4). Cross-sectional imaging revealed no cavities in the pure Ga samples, whereas numerous cavities were observed in MTLM (Fig. 2B). Elemental mapping revealed the spatial distributions of Ga, Fe, and O around microcavities (Fig. 2C). Although vacuum drying removed residual acid, the acid-assisted preparation process remained essential. EGaIn and Fe microparticles mixed without acid treatment were still magnetically responsive, but showed no obvious volumetric growth upon heating (fig. S5). Fe incorporation and heating alone did not induce the volumetric expansion that was observed in MTLM. During material preparation, the Fe content was adjusted to balance magnetic responsiveness with the fluidity required for target delivery (fig. S6).

Thermally induced expansion of internal microcavities contributes to the volumetric expansion of MTLM (fig. S7). We further demonstrated that the stable expanded structure resulted not only from the expansion of pre-existing microcavities but also from heat-induced solidification, which was essential for shape fixation (Fig. 2D). AMF-induced magnetothermal heating produced pronounced volumetric growth, and the expanded structure remained after heating ceased. Reducing the external pressure also caused MTLM to expand, but restoring ambient pressure returned the material toward its initial state. Therefore, the entrapped gas provides a physical capacity for expansion. However, this effect is reversible and cannot explain the persistent structure produced by AMF activation.

The structural solidification of the expanded material is primarily attributed to oxidation during heating. Under AMF activation, the volume of MTLM increased to more than four times its initial value after 20 min of heating (Fig. 2E). Optical tracking of individual cavities showed the same trend at the microscale: cavities enlarged progressively during activation (fig. S8), following a trend similar to that of the increase in bulk volume (movie S1). In an aqueous environment, this growth occurred during AMF heating at temperatures that are relatively mild for human tissues, enabling scaffold formation (Fig. 2F). Without AMF, MTLM in the aqueous environment showed no appreciable spontaneous expansion over the observed period (fig. S9). Water exposure alone was therefore insufficient to produce rapid volume growth.

Chemical analysis provided a second part of the mechanism. $H_2$ was detected during AMF-induced expansion (fig. S10), and its generation synergistically promoted the expansion of pre-existing gas microcavities during thermal activation. Oxidation and oxyhydroxylation of newly exposed gallium interfaces further stabilize the expanded structure (Fig. 2G). Raman spectra showed signals attributable to gallium oxide and gallium oxyhydroxide (Fig. 2H). Elemental analysis showed an increased oxygen fraction after activation (Fig. 2I), while X-ray photoelectron spectroscopy revealed a larger contribution from oxidized Ga species (fig. S11). These findings are consistent with the following reaction:

$$2Ga+4H_2O \rightarrow 2GaOOH+3H_2\uparrow. \quad (1)$$

Related Ga-based liquid-metal foams have also been reported to generate $H_2$ and form GaOOH-rich porous matrices in the presence of water for one day (*13*), supporting the chemical plausibility of this pathway. The complex chemistry of liquid-metal surface oxides has also been recognized as an important factor governing interfacial properties and material behavior (*21*).

For MTLM, the pre-existing gas-filled microcavities provide the initial space for expansion, while $H_2$ generated during activation provides an additional gas source as the cavities grow. At the same time, oxidation and oxyhydroxylation of newly exposed Ga interfaces stabilize the developing porous structure. Therefore, the persistent scaffold appears to arise from the coupling of gas-driven growth with interfacial oxidation and mineralization, rather than from physical gas expansion alone.

Activation also changed how MTLM interacted with wet tissue. Unactivated MTLM was readily displaced from porcine tissue upon inclination, whereas activated MTLM remained in place even after complete inversion (Fig. 2J, K). Cross-sectional imaging showed close contact between the porous structure and the tissue surface (Fig. 2L). The structural transition therefore increased not only bulk mechanical persistence but also retention at a wet tissue interface.

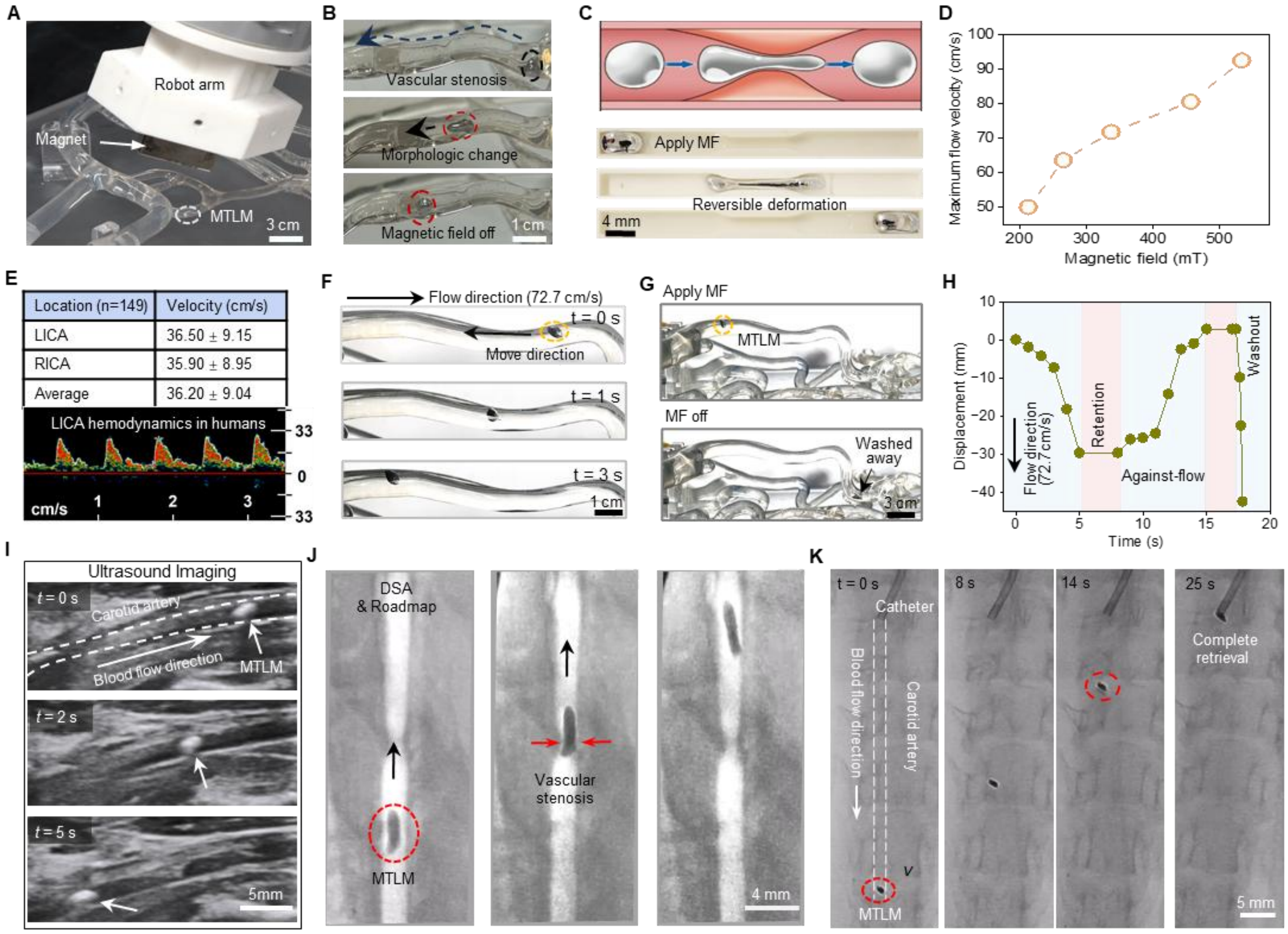

**Fig. 3. Liquid-state MTLM enables magnetic navigation through complex vascular geometries and retention under high flow.**

(**A**) Robotic arm-assisted magnetic navigation of MTLM in a biomimetic vascular model. (**B**) Images of reversible deformation of MTLM during magnetically guided passage through a narrow segment in an *in vitro* vascular model. (**C**) Schematic and experimental images showing axial elongation of MTLM within a constricted channel and recovery toward its original shape after passage. (**D**) Maximum flow velocity permitting magnetic retention of MTLM as a function of magnetic flux density, defined as the highest velocity at which MTLM remained at the target location for 30 seconds without evident downstream displacement. (**E**) Human internal carotid artery blood flow velocities measured by transcranial Doppler (TCD) ultrasonography and a representative Doppler trace. LICA, left internal carotid artery; RICA, right internal carotid artery. Data are mean ± SD (n = 149). (**F**) Representative sequential images showing magnetically driven upstream motion of MTLM at a flow velocity of 72.7 cm $s^{-1}$. (**G**) Magnetic field–dependent retention of MTLM under flow. MTLM remains at the target location while the field is applied and is rapidly washed downstream after field removal. (**H**) Displacement–time trajectory under continuous flow, showing sequential downstream transport, retention at the target location, upstream motion, and washout after field removal. (**I**) Ultrasound images of magnetically guided MTLM motion in a rabbit carotid artery *in vivo*. (**J**) Digital subtraction angiography (DSA) and roadmap images showing MTLM deforming to traverse an experimentally induced vascular stenosis *in vivo*. (**K**) Sequential DSA images showing magnetically guided upstream motion of MTLM against blood flow and subsequent retrieval into a catheter *in vivo*.

### The liquid precursor remains controllable before structural commitment

The MTLM provided intrinsic radiopacity comparable to clinically used embolic materials, enabling fluoroscopic visualization without an additional radiopaque additive (fig. S12). Before remote activation, MTLM remained deformable enough to navigate confined vascular pathways while retaining magnetic control under flow. In its unactivated state, the MTLM could be translated, deformed, and reoriented by an external magnetic field (Fig. 3A and movie S2). It remained maneuverable along curved and closed-loop paths and could be switched among downstream transport (fig. S13A), stationary retention, and upstream motion under continuous flow (movies S3 and S4). Thus, the liquid state enabled not only injection but also controlled transport.

The fluidity also allowed MTLM to adapt to severe geometric confinement. Upon entering a narrow segment, the MTLM elongated, passed through a constriction smaller than its initial diameter, and recovered toward its original morphology after exiting (Fig. 3B). In the channel narrowing from 4 to 2 mm, MTLM traversed the constriction continuously without fragmentation (Fig. 3C and fig. S14). Magnetic guidance further enabled selective delivery to aneurysms at different vascular locations and redistribution among multiple aneurysm targets *in vitro* (figs. S15 and S16 and movie S5).

We then tested whether this mobility could be maintained under substantial flow. The maximum flow velocity that MTLM could resist increased with magnetic flux density, whereas removal of the field led to rapid downstream displacement (Fig. 3D). Doppler measurements gave a mean internal carotid artery velocity of approximately 36.2 cm $s^{-1}$, and patient-derived hemodynamic analysis of 57 bifurcation and sidewall aneurysms showed that velocities at the aneurysm neck and within the sac were generally below those used in our high-flow challenge (Fig. 3E and fig. S17A). In the circulation model, flow velocity was increased to approximately 72.7 cm $s^{-1}$ (Fig. 3F and fig. S17B). Even under this condition, MTLM could be driven upstream, held at a prescribed location, or released into downstream transport by adjusting the magnetic field (Fig. 3, G and H, and movie S6). Positional control before activation was externally imposed and reversible. Magnetic responsiveness was essential for maintaining this control. Under identical flow conditions, MTLM progressively accumulated within the aneurysm under magnetic guidance, whereas pure liquid metal failed to remain localized and was rapidly carried into the parent vessel (movie S7). The two material characteristics thus served different roles: liquid-state deformability accommodated vascular geometry, and magnetic responsiveness opposed hemodynamic displacement.

The same behavior was retained *in vivo* and could be followed by ultrasound and DSA. In rabbit vessels, ultrasound resolved magnetically guided MTLM motion, while DSA visualized changes in position and shape as the precursor traversed a constriction and moved against blood flow (Fig. 3, I to K, and movie S8 and S9). In a canine model, MTLM was also guided from a distant carotid location toward a target aneurysm (fig. S18), extending the navigation experiments from vascular phantoms to intact circulation.

An additional advantage of delaying activation was the opportunity for procedural correction. When MTLM was intentionally displaced into a nontarget vascular region, the material remained visible under angiographic imaging and could be magnetically recaptured and retrieved before activation (Fig. 3K and movie S9). Delivery therefore did not immediately commit the MTLM to an irreversible implant state: its position could still be adjusted, or the material removed, before wireless activation. This correction window contrasts with conventional liquid embolic systems, in which repositioning becomes progressively more difficult once precipitation, polymerization, or gelation has begun.

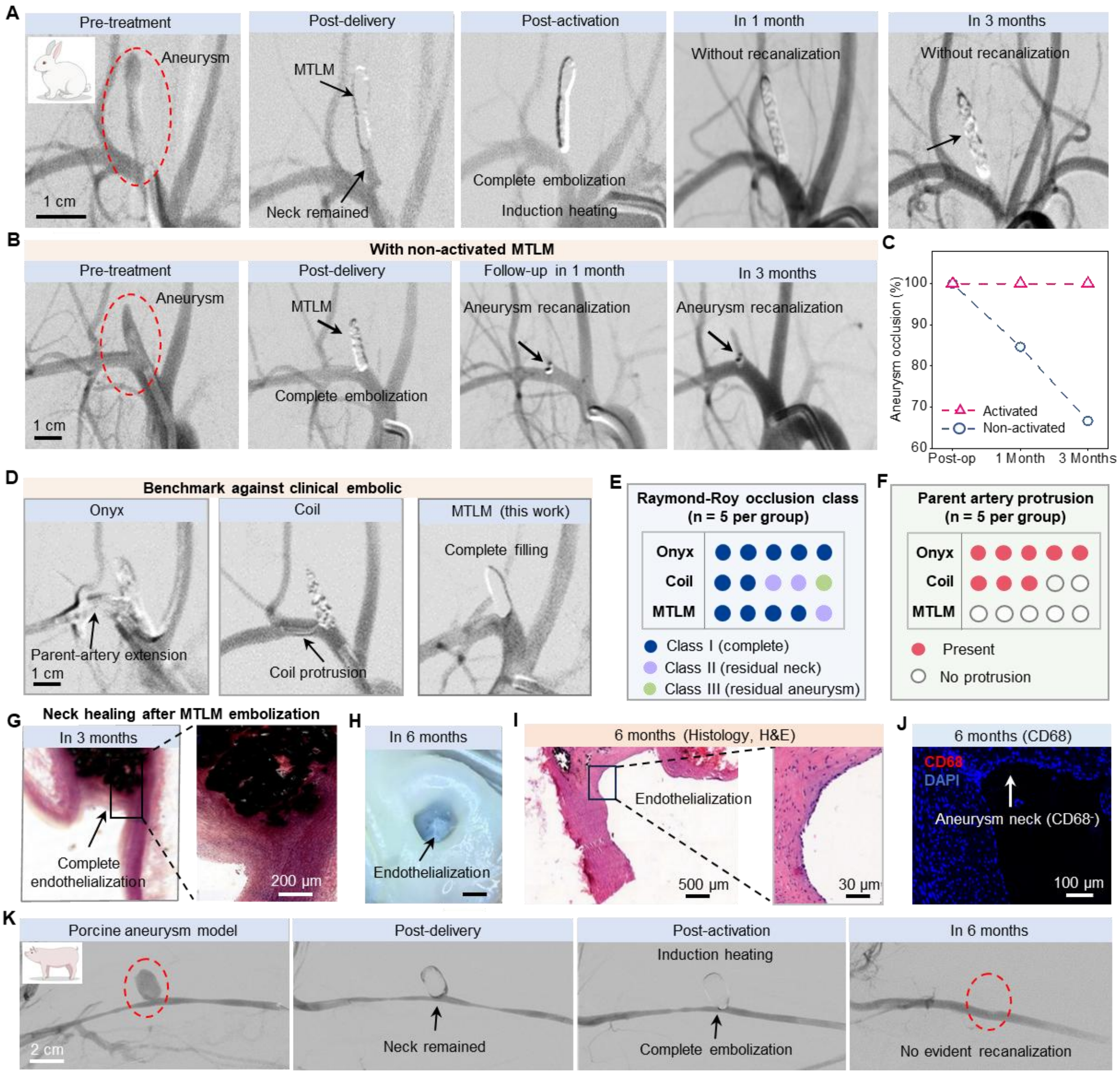

**Fig. 4. Postdeployment activation enables durable aneurysm occlusion with parent-artery preservation.**

(**A**) DSA images showing sustained aneurysm occlusion at 1 and 3 months after MTLM delivery and AMF activation. (**B**) DSA images of a rabbit aneurysm treated with MTLM without AMF activation. Substantial initial filling is followed by progressive recanalization at 1 and 3 months. (**C**) Aneurysm occlusion over time with and without AMF activation, expressed as the percentage of aneurysm volume excluded from contrast opacification. (**D**) DSA images after embolization with Onyx, coils, or MTLM, showing Onyx extension into the parent artery, coil protrusion at the aneurysm neck, and MTLM conformally filled the aneurysm sac. (**E**) Raymond–Roy occlusion grades for aneurysms treated with Onyx, coils, or MTLM (n = 5 aneurysms per group). (**F**) Proportion of aneurysms with embolic material extending or protruding into the parent artery in each treatment group. No parent-artery protrusion was observed in the MTLM group. (**G**) Histological images at 3 months after AMF-activated MTLM embolization, showing continuous neotissue coverage across the aneurysm neck. (**H**) Gross views of the aneurysm neck at 6 months after MTLM embolization, showing continuous tissue coverage after treatment. Scale bar, 1 mm. (**I**) Hematoxylin and eosin (H&E) staining at 6 months, showing continuous neointima-like tissue across the aneurysm neck. (**J**) CD68 immunofluorescence with DAPI nuclear counterstaining at 6 months, showing no evident accumulation of $CD68^+$ macrophages at the aneurysm neck. (**K**) DSA images of MTLM embolization in a porcine aneurysm model, showing the untreated aneurysm, residual neck filling after initial delivery, complete occlusion after AMF activation, and sustained occlusion without evident recanalization at 6 months.

## Remote activation converts transient filling into durable vascular reconstruction

Activated MTLM showed lower mass-normalized magnetization (fig. S19), remained anchored in a vascular model after activation (fig. S20), and sealed a blood-column model for 10 days without visible leakage (fig. S21). We next examined whether postdeployment activation was necessary for durable embolization in a rabbit aneurysm model. MTLM activated by AMF after deployment formed a persistent porous structure in situ and maintained aneurysm occlusion at 1 and 3 months (Fig. 4A). By contrast, unactivated MTLM initially filled the aneurysm sac but showed declining occlusion during follow-up, with recanalization evident at both time points (Fig. 4B). Although the liquid precursor initially filled the aneurysm, sustained occlusion required subsequent activation (Fig. 4C). In an *ex vivo* aneurysm model, AMF-activated MTLM remained anchored under a fluid challenge that washed out nonactivated MTLM (fig. S22). DSA parametric maps illustrated changes in intra-aneurysmal contrast transit (fig. S23), and computational fluid dynamics simulations showed reduced intra-aneurysmal velocity and wall shear stress under the modeled conditions (figs. S24 to S26).

We then compared MTLM with representative liquid and solid embolic approaches in a wide-neck aneurysm model. MTLM embolization was technically completed in all five rabbits. Onyx filled the aneurysm but extended into the parent artery in all five animals, whereas detachable coils could leave residual neck filling or protrude into the parent vessel. MTLM conformed to the aneurysm geometry in the liquid state and, after activation, formed a stable structure while preserving parent-artery patency (Fig. 4D). All five Onyx-treated aneurysms were classified as Raymond–Roy class I. In the coil group, two aneurysms were class I, two were class II, and one was class III. Four of five MTLM-treated aneurysms achieved class I occlusion and one was class II, with no parent-artery protrusion observed (Fig. 4E, F).

Neurological and feeding assessments revealed differences among the treatment groups. In this cohort, MTLM-treated rabbits showed fewer observable neurological deficits and better-preserved food intake than Onyx-treated rabbits (fig. S27). These outcomes accompanied a high degree of aneurysm occlusion and preserved parent-vessel patency under the conditions tested. MTLM therefore combined a high degree of aneurysm occlusion with preserved parent-vessel patency and fewer observable neurological deficits than Onyx treatment.

Long-term follow-up showed that persistent occlusion was accompanied by progressive healing across the aneurysm neck. Rabbit aneurysms remained occluded at 3 months and showed no evident recanalization at 6 months. Gross examination and histological analysis revealed progressive tissue coverage across the neck, together with collagen-rich remodeling and maturation of the repaired region (Fig. 4, G to I, and fig. S28 and S29). CD68 immunostaining at 6 months showed limited macrophage accumulation at the aneurysm neck (Fig. 4J). The activated scaffold therefore remained compatible with continued tissue remodeling during long-term vascular exclusion.

We further evaluated durability in a porcine aneurysm model. MTLM was delivered into the aneurysm and activated in situ, producing complete angiographic occlusion that remained stable without evident recanalization at 6 months (Fig. 4K). The approach was also tested in a porcine cerebrovascular malformation model. MTLM delivery reduced lesion opacification, and subsequent AMF activation eliminated the residual abnormal angiographic signal (fig. S30), extending the approach from focal aneurysm cavities to a more distributed vascular lesion.

Cytocompatibility and hemocompatibility were assessed using cell viability, live/dead staining, hemolysis, and platelet-consumption assays (figs. S31 to S33). Hematological, coagulation, hepatic, and renal measurements were obtained during follow-up (fig. S34). Circulating Ga and In were reported below the detection limits at the measured time points (table S1). Histological examination of the sampled organs and vascular tissues showed no overt pathological abnormalities in the representative sections (fig. S35).

Additional studies explored possible extensions of the platform. Incorporation of sodium alginate enabled thrombin loading and release (fig. S36), while combining aneurysm volume with MTLM expansion kinetics was explored as a basis for selecting the initial material dose and activation time (fig. S37 and fig. S38). Comparison with representative intravascular strategies further highlighted a combination of capabilities rarely available within a single material system, including liquid-state conformability, magnetic positioning under flow, preactivation retrievability, and postdelivery generation of additional therapeutic volume and a persistent scaffold (table S2).

## Discussion

Endovascular embolization places different demands on a material at different stages of treatment. Procedure-related vessel or aneurysm rupture and post-treatment recurrence remain major challenges in endovascular aneurysm therapy (*9, 22, 23*). During delivery, the material must pass through narrow and tortuous vascular pathways; after implantation, it must resist flow and remain mechanically stable within a much larger lesion. MTLM addresses these requirements sequentially rather than within a single fixed material state. It remains deformable and magnetically controllable during transport, while its final porous scaffold is generated only after localization. In this sense, the therapeutic structure is not carried through the catheter in its final form but constructed at the target site.

Conventional coil embolization relies on coil packing and subsequent thrombus formation, with thrombus remodeling and coil compaction potentially contributing to recanalization (*9, 24*). MTLM instead expands and mechanically stabilizes in situ, providing direct structural support through the transformed material. This transition is distinct from simple curing, swelling, or shape recovery. MTLM is prestructured during preparation but remains fluid before activation. Magnetothermal heating drives volumetric growth while newly exposed Ga interfaces undergo oxidation and oxyhydroxylation, stabilizing the emerging porous architecture as it forms. Thus, the scaffold is constructed during activation, with its volume and internal architecture developing after delivery rather than being predetermined in a pre-existing implant.

Coil packing offers limited control over the final three-dimensional arrangement and may require repeated adjustments to accommodate aneurysm geometry (*25, 26*). The fluid MTLM precursor conforms to the aneurysm sac while allowing magnetic control of its distribution before fixation, which may reduce the need for repeated adjustments. This fluid state also permits repositioning under flow, retrieval after off-target displacement, and additional delivery when filling is incomplete. After activation, MTLM forms a persistent scaffold that maintains aneurysm exclusion and supports tissue remodeling. Progressive neck coverage and limited macrophage accumulation during long-term follow-up suggest compatibility with vascular healing beyond acute occlusion.

Several questions remain before translation. Although processed MTLM remained stable in aqueous environments over the tested period without activation, longer residence times and a broader range of physiological conditions will need to be evaluated. The extent to which volumetric growth slows or terminates after withdrawal of AMF also requires quantitative definition, particularly under vascular confinement. Magnetothermal activation must be characterized more fully in deeper and smaller vessels, including spatial temperature gradients, local heating around magnetic inclusions, and performance under continuous blood flow (*27*). Finally, the long-term mechanical evolution and biological fate of the mineralized scaffold beyond the present follow-up period remain unknown. These factors will ultimately define the operating window and safety margins needed for clinical translation.

In conclusion, we developed a remotely activated magnetothermal liquid implant that remains deformable and controllable during endovascular delivery but forms a mechanically persistent porous scaffold only after target localization. Magnetothermal activation couples volumetric growth with interfacial mineralization, enabling durable aneurysm occlusion and subsequent healing. More broadly, this work introduces postdelivery construction as a strategy in which a small, controllable precursor generates additional therapeutic volume and a mechanically persistent architecture only after reaching the target site.

**Ethics statement**: The retrospective human hemodynamic study was approved by the Ethics Committee of Beijing Tiantan Hospital (KY2016-023-01); All animal procedures were approved by the Institutional Animal Care and Use Committee of Beijing Tonghe Litai Biotechnology Co., Ltd. (IACUC-D2024019).

**Supplementary Materials**

Materials and Methods

Figs. S1 to S38

Tables S1 to S3

References (28–45)

Movies S1 to S9